\documentclass[]{spie}  

\usepackage{amsmath,amsfonts,amssymb}
\usepackage{graphicx}
\usepackage[hang,labelfont=it]{caption}
\usepackage{subcaption}
\usepackage[colorlinks=true, allcolors=black]{hyperref}

\title{A metrological characterization of the SPEED test-bed PIAACMC components}

\author[a]{K. Barjot}
\author[b]{P. Martinez}
\author[b]{M. Beaulieu}
\author[b]{C. Gouvret}
\author[b]{A. Marcotto}
\author[c]{O. Guyon}
\author[d]{M. Belhadi}
\author[d]{A. Caillat}
\author[d]{T. Behaghel}
\author[e]{J.M. Le Duigou}
\author[d]{K. Dohlen}
\author[d]{A. Vigan}

\affil[a]{LESIA, Observatoire de Paris, Universit\'e PSL, CNRS, Sorbonne Universit\'e, Universit\'e de Paris, 5 place Jules Janssen, 92195 Meudon, France}
\affil[b]{Universit\'e C\^ote d'Azur, Observatoire de la C\^ote d'Azur, CNRS, Laboratoire Lagrange, France}
\affil[c]{National Astronomical Observatory of Japan, Subaru Telescope, 650 North Aohoku Place, Hilo, HI 96720, U.S.A.}
\affil[d]{Aix Marseille Univ., CNRS, CNES, LAM, Marseille, France}
\affil[e]{Centre National d'\'Etudes Spatiales, 18 avenue Edouard Belin 31401 Toulouse cedex 9, FRANCE}

\authorinfo{Send correspondence to K. Barjot, E-mail: kevin.barjot@obspm.fr}

\begin{document} 
\maketitle

\begin{abstract}
  The segmented pupil experiment for exoplanet detection (SPEED) facility aims to improve knowledge and insight into various areas required for gearing up high-contrast imaging instruments adapted to the unprecedented high angular resolution and complexity of the forthcoming extremely large telescopes (ELTs).
  SPEED combines an ELT simulator, cophasing optics, wavefront control and shaping with a multi-deformable mirror (DM) system, and optimized small inner-working angle (IWA) coronagraphy. The fundamental objective of the SPEED setup is to demonstrate deep contrast into a dark hole optimized for small field of view and very small IWA, adapted to the hunt of exoplanets in the habitable zone around late-type stars. 
  SPEED is designed to implement an optimized small IWA coronagraph: the phase-induced amplitude apodization complex mask coronagraph (PIAACMC). The PIAACMC consists in a multi-zone phase-shifting focal plane mask (FPM) and two apodization mirrors (PIAA-M1 and PIAA-M2), with strong manufacturing specifications. Recently, a first-generation prototype of a PIAACMC optimized for the SPEED facility has been designed and manufactured. The manufacturing components exhibit high optical quality that meets specifications. 
  In this paper, we present how these components have been characterized by a metrological instrument, an interferential microscope, and then we show what is yielded from this characterization for the FPM and the mirrors. Eventually, we discuss the results and the perspectives of the implementation of the PIAACMC components on the SPEED setup.
\end{abstract}


\section{Description}

Since the first direct image of an exoplanet \cite{chauvin2004giant} sixteen years ago, only several tens of exoplanets have been directly imaged being mostly giant Jupiter-like or very young planets in orbits far from their host stars. Hence, in order to widen the variety of exoplanets down to Earth-like objects in small orbits, it is crucial to develop an instrument capable of high contrast imaging at low inner-working angle (IWA). The forthcoming ELTs will allow observations at unprecedented high resolution but they will add new technical challenges to overcome, because of their inherent complexity (e.g. the segmentation of the primary mirror).

The segmented pupil experiment for exoplanet detection (SPEED) is a testbed aiming at addressing these new challenges, for direct imaging of exoplanets at small IWA with the light coming from a complex pupil such as the future ESO ELT. To do so, SPEED uses a segmented telescope simulator and is composed of (1) a cophasing part (working in the visible wavelengths) and (2) a high contrast part (working in the near-infrared wavelengths). High contrast imaging is obtained with a PIAACMC \cite{guyon2003phase} (phase-induced amplitude apodization complex mask coronagraph) and two deformable mirrors for wavefront control and shaping.

The PIAACMC \cite{guyon2010high} designed for the SPEED project is made of three optical parts: (1) a pair of aspheric apodization mirrors (named M1 for the first one on the optical path and M2 for the next one), (2) a reflective complex focal plane mask (FPM), and (3) a Lyot stop that blocks diffracted light.

\begin{figure}[!htbp]
    \begin{subfigure}{.25\textwidth}
        \includegraphics[width=0.95\linewidth]{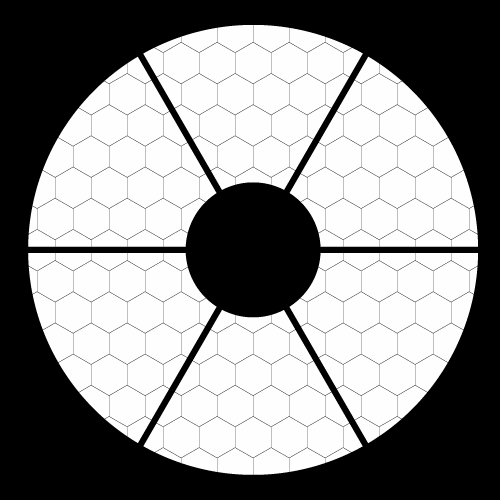}
    \end{subfigure}%
    \begin{subfigure}{.25\textwidth}
        \includegraphics[width=0.95\linewidth]{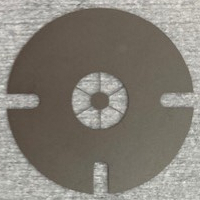}
    \end{subfigure}%
    \begin{subfigure}{.25\textwidth}
        \includegraphics[width=0.95\linewidth]{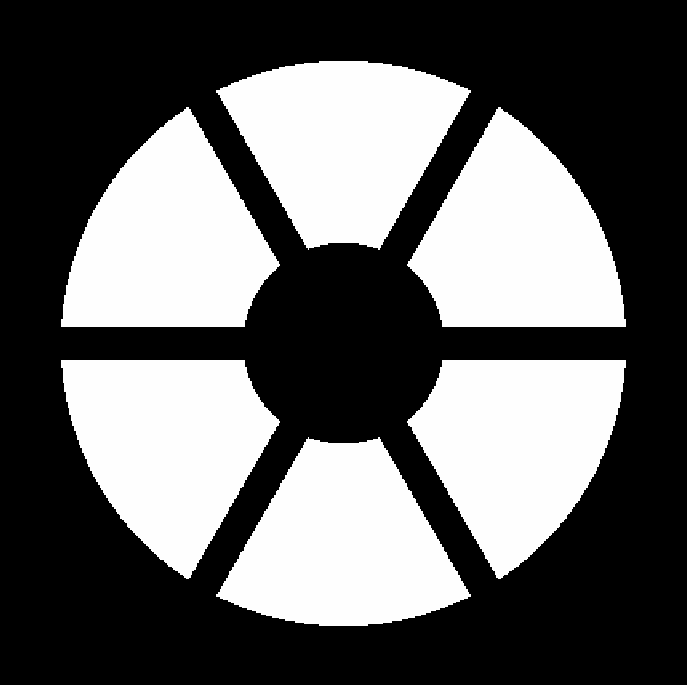}
    \end{subfigure}%
    \begin{subfigure}{.25\textwidth}
        \includegraphics[width=0.95\linewidth]{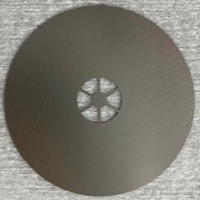}
    \end{subfigure}
    \caption{From left to right the simulated SPEED pupil, the physical mask mimicking the ESO central obscuration and spider vanes, the numerical map of the PIAACMC Lyot stop, and the physical Lyot stop installed on the bench.}
    \label{fig:SPEEDPupilLyot}
\end{figure}

The first apodization mirror M1 compresses the light beam while the second mirror M2 corrects for the optical path length errors introduced by M1. As a result, the light intensity function becomes a prolate-like function such that the stellar light focused onto the FPM is efficiently decreased via destructive interferences and diffracted out of the optical axis. Finally, the Lyot stop is a mask designed to block this latest rejected light and the light diffracted by the spider of the telescope. The figure~\ref{fig:SPEEDPupilLyot} presents from left to right: the simulated SPEED pupil, the physical mask mimicking the ESO central obscuration and spider vanes (the pupil segmentation is introduced with an IRIS AO deformable mirror), the numerical map of the Lyot stop, and the physical Lyot stop installed on the bench. These components are manufactured by laser cutting within a few micron constraint.

\section{Components and measurement methods}

\subsection{Component introduction and specifications}

\begin{figure}[!htbp]
    \begin{subfigure}{.34\textwidth}
        \includegraphics[width=0.9\linewidth]{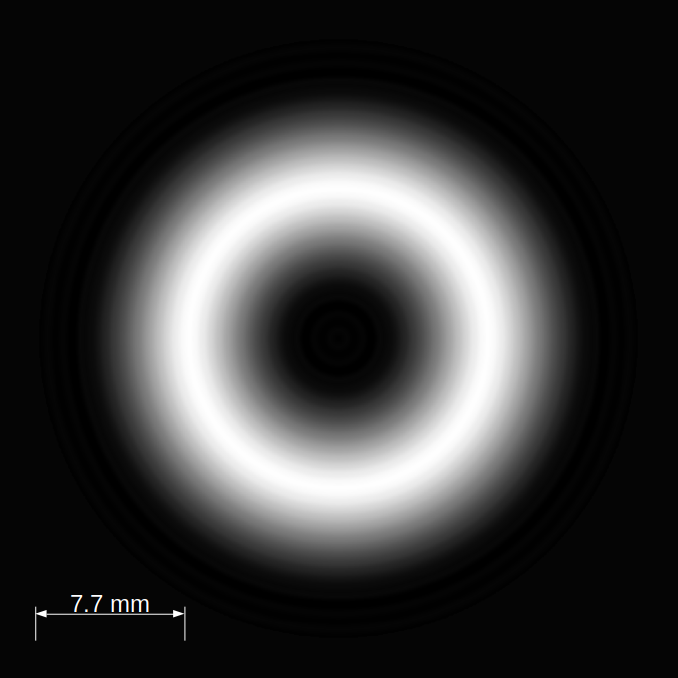}
    \end{subfigure}%
    \begin{subfigure}{.34\textwidth}
        \includegraphics[width=0.9\linewidth]{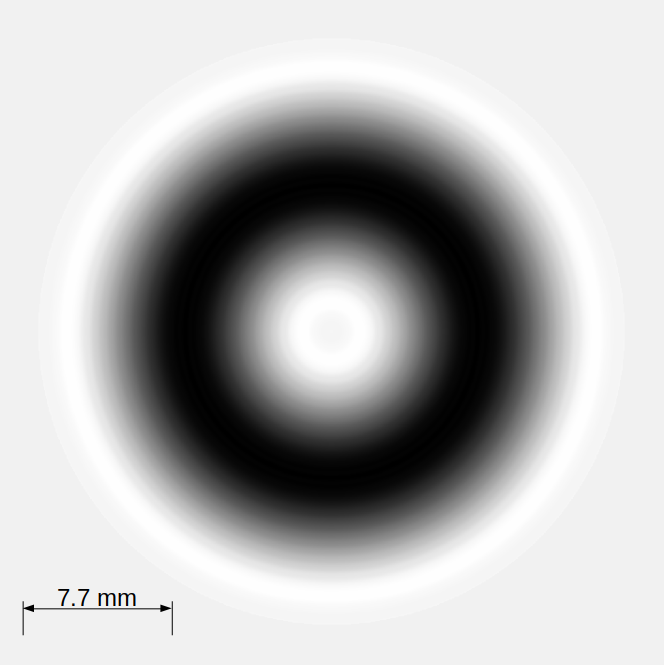}
    \end{subfigure}%
    \begin{subfigure}{.34\textwidth}
        \includegraphics[width=0.9\linewidth]{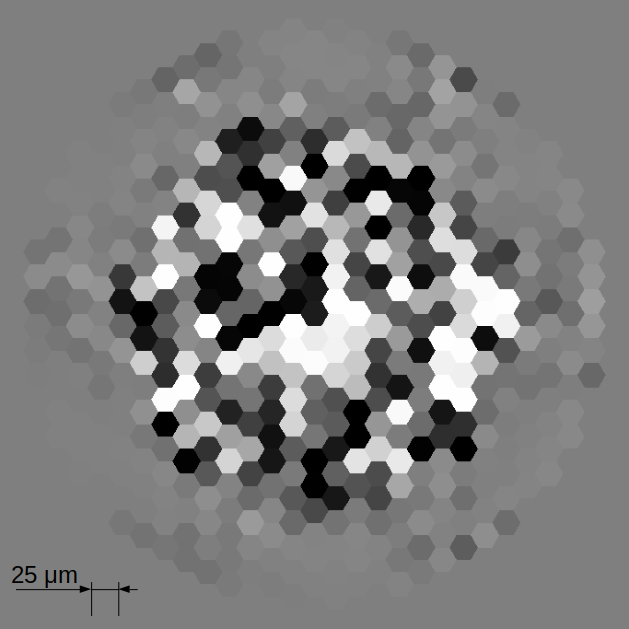}
    \end{subfigure}
    \caption{2D surface map: M1 (left), M2 (center), FPM (right)}
    \label{fig:specProfile}
\end{figure}

SILIOS technologies manufactured the two apodization mirrors and the FPM. The surface profiles was realized by an engraving and photolithographic process dictated by the specified profiles \cite{martinez2018design} shown in figure~\ref{fig:specProfile}, on the left and middle for the two mirrors and the right for the FPM. The two mirror profiles are defined by discretized concentric rings at different depths to ensure a centrally-symmetric profile. The FPM surface is a pavement of $499$ hexagons at determined depths. They were made into fused silica substrates with a diameter of $525 \, \mu m$ with a reflective gold coat.

The studied features of this paper are shown in table~\ref{tab:specifications}. For every feature reported in the first column, its manufacturing specification is described in the second column for the PIAA mirrors and in the third column for the FPM. In the next sections, we will show the metrological characterization study of these features for the SPEED testbed components.

\begin{table}[!htbp]
    \centering
    \begin{tabular}{|c|c|c|}
        \hline
        Features & PIAA mirrors & FPM \\
        \hline
        Useful area & $7.7 \, mm$ & $525 \, \mu m$ \\
        \hline
        Sag & $\sim 4 \, \mu m$ & $\pm 0.4 \, \mu m$ \\
        \hline
        Sag local deviation tolerance (rms) & $<50 \, nm$ (best effort $<25 \, nm$) & $<5 \, nm$ (best effort $<3 \, nm$) \\
        \hline
        Number of hexagon & n/a & $499$ \\
        \hline
        Hexagon dimension & n/a & $25 \, \mu m$ \\
        \hline
        Hexagon dimension tolerance & n/a & $<0.5 \, \mu m$ \\
        \hline
    \end{tabular}
    \caption{Specifications of the manufactured components features.}
    \label{tab:specifications}
\end{table}

\subsection{Optical metrology tools}

To conduct the metrological characterization, we benefited from the equipment of the \textit{Laboratoire d'Astrophysique de Marseille} (LAM). We used the WYKO NT9100 \cite{FicheTechniqueWYKO} then referred as \textit{WYKO}, which is an interferential microscope. It uses a LED light split into two beams reflected on a calibrated mirror and reflected on the sampled surface. In the way of a Michelson interferometer, the combination of these two reflected beams is measured in order to infer the sampled surface depth.

Moreover, since the FPM has not the same dimensions ($525 \, \mu m$) as the mirrors ($7.7 \, mm$), the field of views of the measurements are adapted: $762 \, \mu m \times  695 \, \mu m$ with a spatial resolution of $1.98 \, \mu m$ for the FPM and $47 \, \mu m \times 63 \, \mu m$ of field of view with a spatial resolution of $97.92 \, nm$ for the mirrors. Hence, the measurements of the mirror surfaces are not fully covered and several images of different regions need to be recombined to obtain the full radial profile of these components. As we will see later, this technique is subject to errors in the reconstruction of the radial profile.

\section{FPM characterization}

\subsection{Manufacturing sizing and cosmetic aspects}
\label{sec:CosmeticAspects}

The diameter of hexagon patterns on the component (such as every tile or the overall tiling), is defined as the diameter of their circumscribed circle. Therefore, using the \textit{WYKO} instrument, we measure the overall pavement diameter to $531 \pm 1 \, \mu m$ (specified to $525 \, \mu m$), which is a discrepancy less than half a single hexagonal segment. In addition, the typical measured diameter of segments is $24.51 \pm 0.07 \, \mu m$ (specified to $25 \pm 0.5 \, \mu m$).

\begin{figure}[!htbp]
    \begin{subfigure}{.5\textwidth}
        \includegraphics[width=0.9\linewidth]{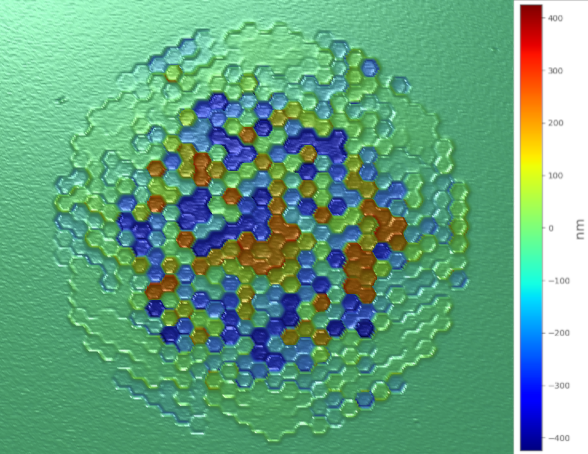}
    \end{subfigure}%
    \begin{subfigure}{.5\textwidth}
        \includegraphics[width=\linewidth]{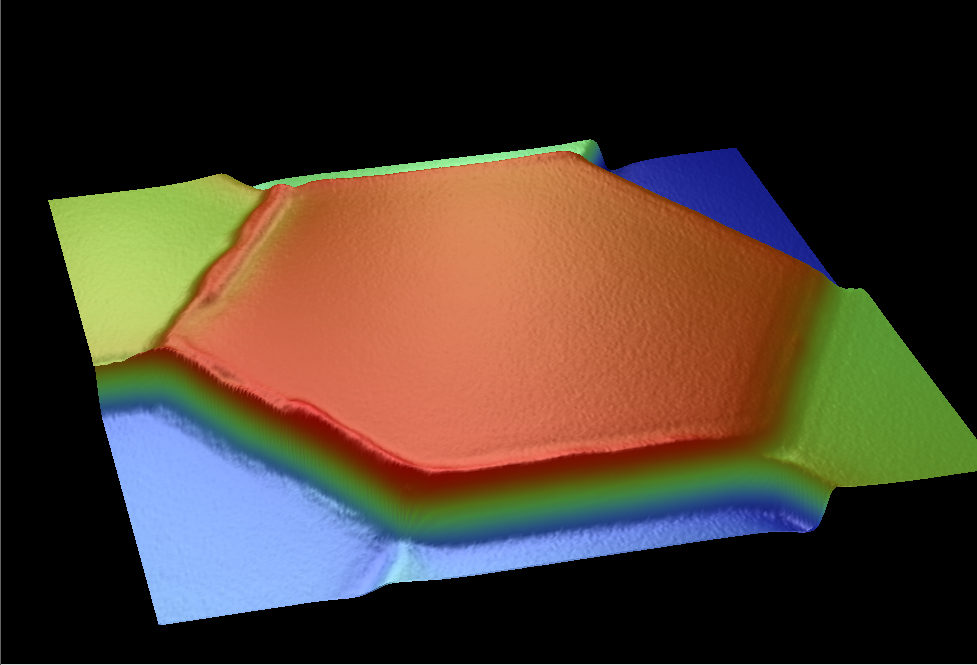}
    \end{subfigure}
    \caption{FPM surface measurements with the \textit{WYKO}: magnification of $\times 5$ on the left and of $\times 100$ on the right.}
    \label{fig:FPMWYKOSurfaceMeas}
\end{figure}

Furthermore, a global cosmetic inspection shows a few number of artifacts on the components originated from manufacturing defaults and contamination, located most of the time outside of the useful area. Their sizes are found relatively small compared to the overall tiling, hence are expected to have no impact on the component performances. Figure~\ref{fig:FPMWYKOSurfaceMeas} shows a FPM global measurement map (on the left) and a higher magnification measurement ($\times 100$ on the right) enlightening a high quality of the segment surface. Besides, on this right image of a small scale measurement, we can observe irregularities on the transitions between hexagonal segments. This edge effect can be explained either by the photolithographic process which builds the phase mask that could leave extra material or by a measurement artifact inherent in interferential microscopy (called \textit{batwings}) which was previously simulated and measured\cite{lehmann2015signal}. On the transition of two parts at different depths, this instrumental error is featured by an extra material in the highest part and a lack of material in the lowest part of the edge. It appears on the FPM measurements and as we will see in section~\ref{sec:MDiscretization} in the mirror measurements. As a result, we observe a satisfying cosmetic quality of the surface of the FPM as well as for the sizing.

\subsection{Engraving measurements}
\label{sec:EngravingMeasurements}

\begin{figure}[!htbp]
    \begin{subfigure}{.5\textwidth}
        \hspace{-1.5cm}
        \includegraphics[width=1.38\linewidth]{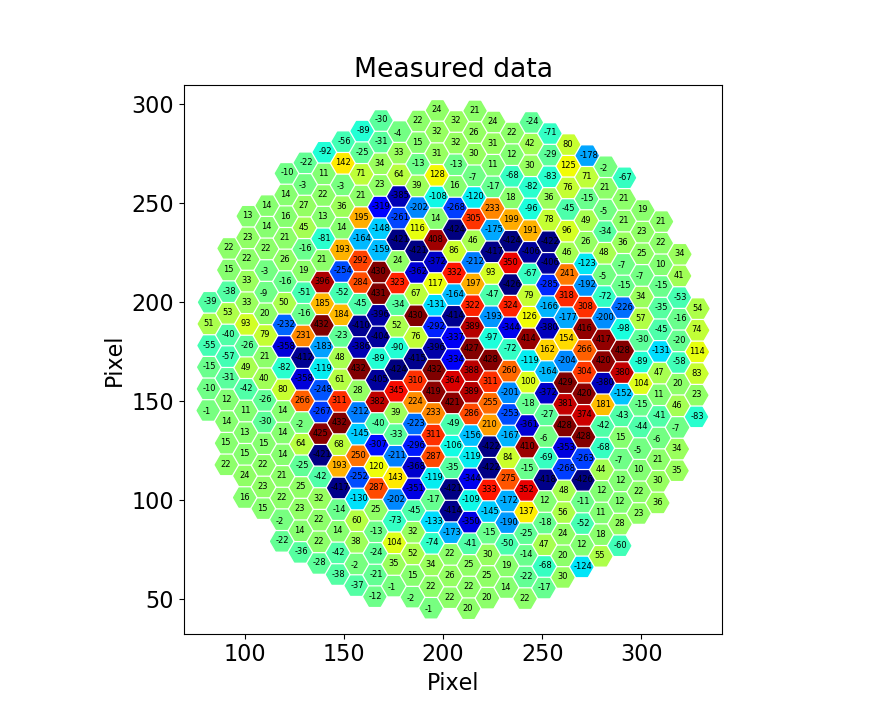}
    \end{subfigure}%
    \begin{subfigure}{.5\textwidth}
        \hspace{-0.4cm}
        \includegraphics[width=0.92\linewidth]{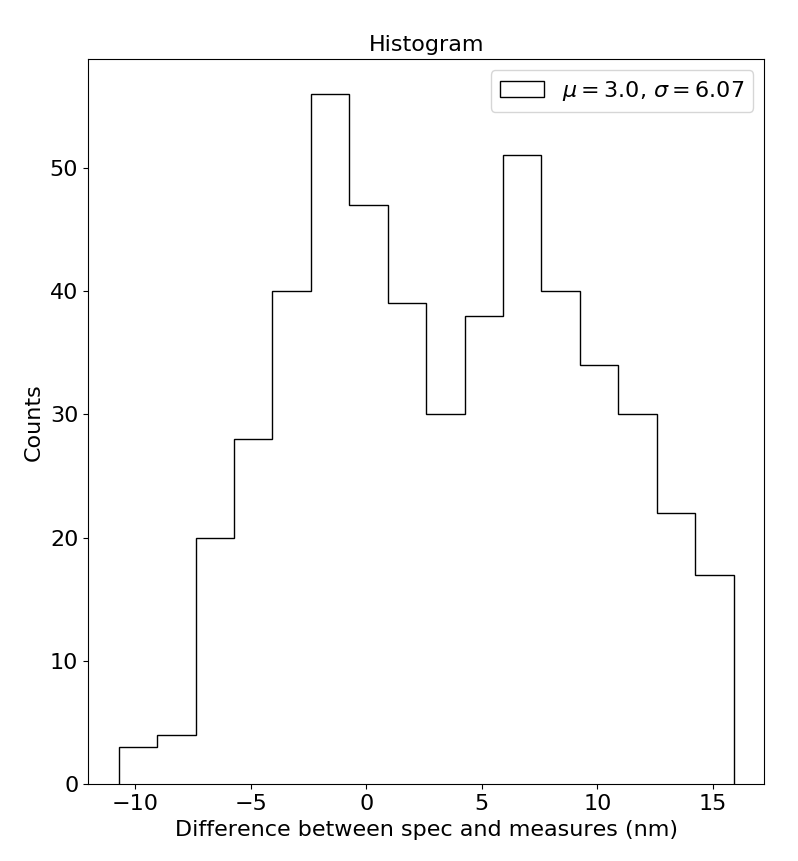}
    \end{subfigure}
    \caption{On the left: reconstructed map of the FPM pavement depths with the measurements of the \textit{WYKO}. \\ On the right: histogram of the difference between the specified sags and the measured sags.}
    \label{fig:SagAnalysis}
\end{figure}

The other crucial feature in the analysis of the FPM is its sag map. Using global measurements (with low magnification of $\times 5$) of the component with the \textit{WYKO}, we can recover the engraved depth of the $499$ hexagons. A map is reconstructed with these measurements and shown on the left of the figure~\ref{fig:SagAnalysis} where the measured sags are color-coded from blue to red from lowest depth to highest depth segments. On the right of the figure~\ref{fig:SagAnalysis} is shown the histogram of the difference between the specification and the measured segment sags. A difference of $3 \, nm \, rms$ in average with a standard deviation of $6.07 \, nm \, rms$ is measured in a range of $-10 \, nm$ and $15 \, nm$, which is found in agreement with the specification of $5 \, nm \, rms$.

The analysis also compares the specifications and the measurements of several specific segments, chosen as following: 
\begin{itemize}
    \item[$-$] the minimal depth segments (specified to $- 424 \, nm$), are measured from $- 421.1 \pm 0.4 \, nm$ to $- 421.9 \pm 0.4 \, nm$,
    \item[$-$] the maximal depth segments (specified to $425 \, nm$) are measured from $427.3 \pm 0.4 \, nm$ to $432.8 \pm 0.4 \, nm$,
    \item[$-$] the adjacent segments presenting the smallest depth difference (specified to $2 \, nm$) is measured $3.45 \pm 0.06 \, nm$
    \item[$-$] the adjacent segments presenting the highest depth difference (specified to $844 \, nm$) is measured $846.2 \pm 0.4 \, nm$.
\end{itemize}

These measurements show small variations of the segments sags compared to the prescribed ones of the order of few nanometers.

Eventually, the surface roughness is measured computing the mean difference between the depth values and the mean depth over a given surface. This surface of interest is defined by a circle over a segment such that the edges are not taken into account, resulting in a measured surface area of $16 \, \mu m$ in diameter (this area is used to build a numerical map of the surface of the component, see section~\ref{sec:Prospective}). The roughness is measured at $2.81 \, nm$ (standard deviation of $1.57 \, nm$). A good quality of the FPM surface roughness is found as also illustrated in the right of figure~\ref{fig:FPMWYKOSurfaceMeas}.

\section{Mirror characterization}

\subsection{Cosmetic aspects}

Similarly to the FPM study, the surface of the mirrors is inspected seeking artifacts or impurities. These are identified as dust, hollows or scratches. They are first identified and localized during a low magnification inspection ($\times 2.75$) then are characterized at higher magnification ($\times 100$). 

\begin{figure}[!htbp]
    \begin{subfigure}{.5\textwidth}
        \hspace{-0.09cm}
        \includegraphics[width=\linewidth]{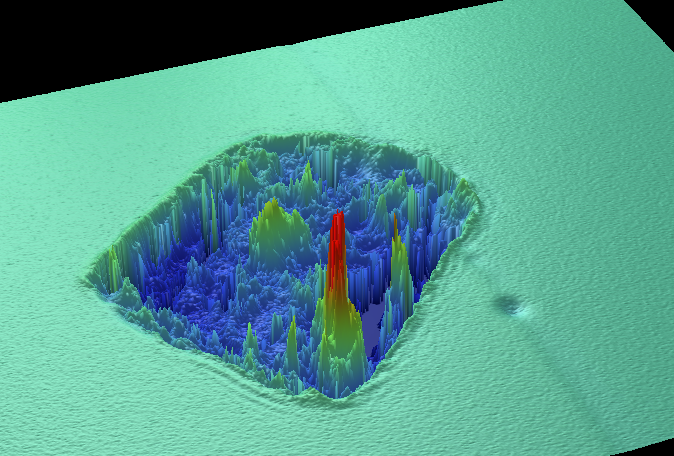}
    \end{subfigure}%
    \begin{subfigure}{.5\textwidth}
        \hspace{0.09cm}
        \includegraphics[width=\linewidth]{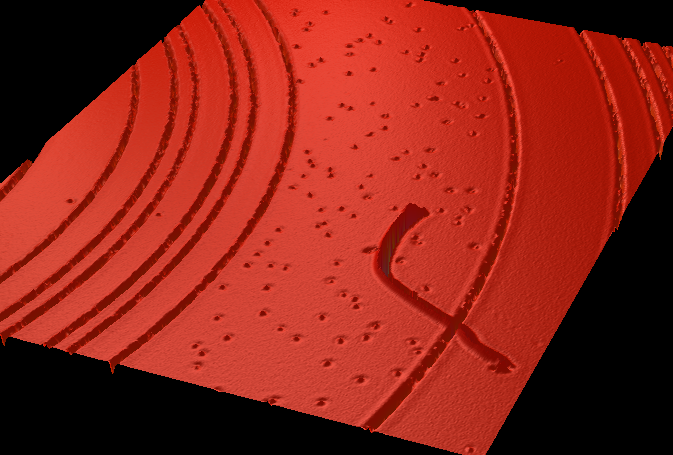}
    \end{subfigure}
    \caption{Mirror artifacts images. On the left: one big impact, on M1 mirror, of size $24.96 \pm 0.03 \, \mu m$ and $\sim 6 \, \mu m$ in depth, located inside the obstruction. On the right: \textit{bubbles} found inside a ring and a fiber-like impurity, both inside the obstruction, on the M2 mirror.}
    \label{fig:MirrorArtifacts}
\end{figure}

It is found that the biggest impurities are located in the obstructed zone which corresponds to the shadow of the secondary mirror of the SPEED pupil, sizing at $30 \, \%$ of the mirror diameters ($7.7 \, mm$). It means that these defects will have no significant impact on the coronagraphic performances. The M1 and M2 biggest artifacts found have a size of respectively $29 \pm 1 \, \mu m$ and $151.6 \pm 0.3 \, \mu m$. They are in both cases inside the obstructed area of the mirrors. As an illustration, the left image of the figure~\ref{fig:MirrorArtifacts} presents one of these artifacts on the M1 mirror with a size of $24.96 \pm 0.03 \, \mu m$ and a depth of $\sim 6 \, \mu m$, also located inside the obstruction. Besides, the most seen artifacts are found with a common size of about $30 \, \mu m$ on both mirrors.

In addition, an impurity identified as a fiber is found on the M2 inside the obstruction, and is visible on the right image of the figure~\ref{fig:MirrorArtifacts}. It has a measured size of $235 \pm 1 \, \mu m$ in length.

Finally, on the M2 mirror only, holes are found looking like \textit{bubbles} as shown in the right of the figure~\ref{fig:MirrorArtifacts}. They are located on a specific ring between $631 \, \mu m$ and $821 \, \mu m$, inside the obstruction. Since they are seen also on spare M2 mirrors (not discussed in the paper), on the same ring, it appears that they are inherent from the manufacturing process of the M2. Their sizes are measured from $1.18 \pm 0.03 \, \mu m$ to $1.47 \pm 0.03 \, \mu m$ in diameter and with $1.3 \, \mu m$ in depth.

\subsection{Radial profiles}
\label{sec:RadialProfile}

\begin{figure}[!h]
    \begin{subfigure}{.5\textwidth}
        \includegraphics[width=\linewidth]{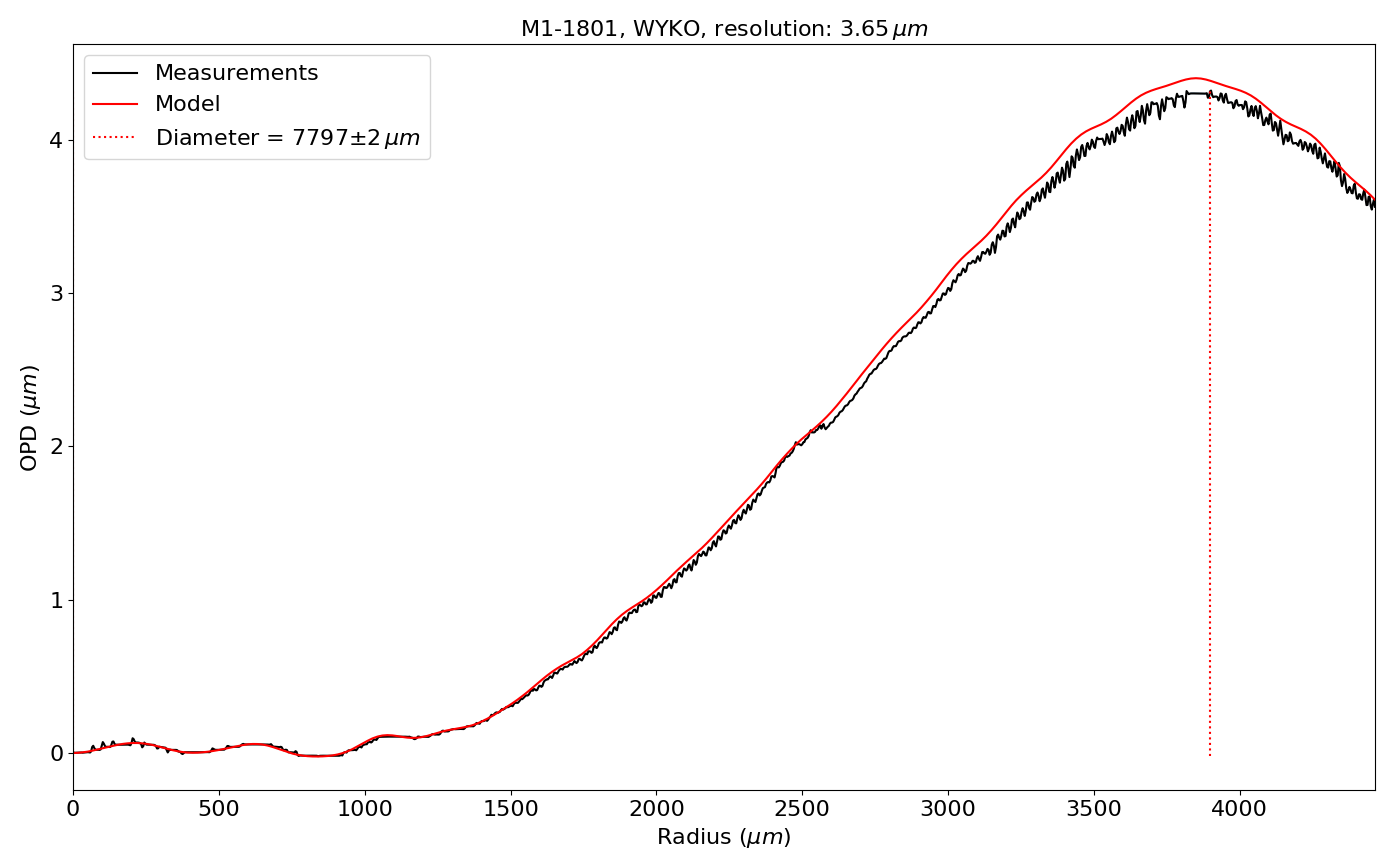}
    \end{subfigure}%
    \begin{subfigure}{.5\textwidth}
        \includegraphics[width=\linewidth]{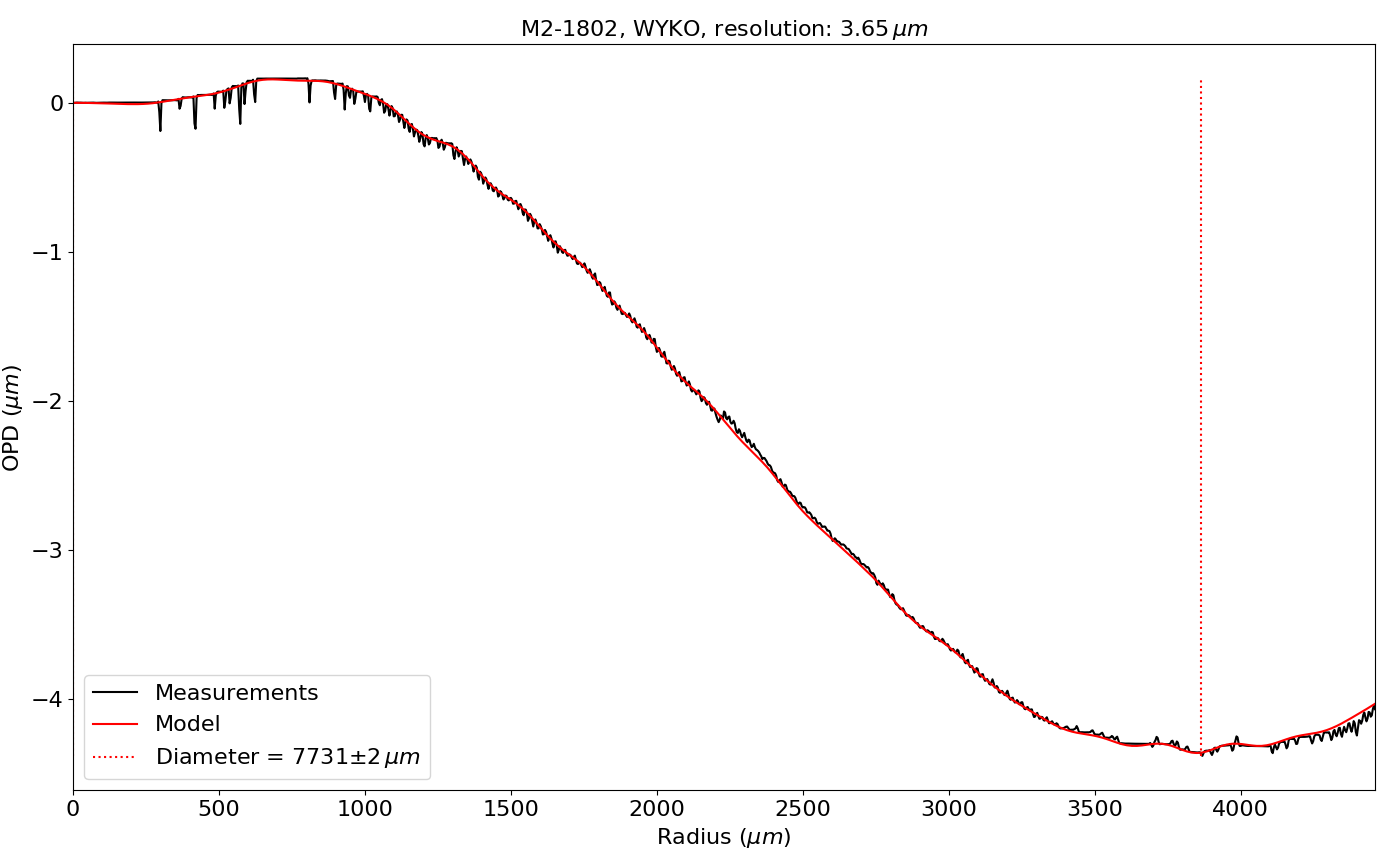}
    \end{subfigure}
    \caption{Radial profiles of the M1 mirror (on the left) and the M2 mirror (on the right) taken from the recombination of several field of view measurements (in black). The prescribed profiles are shown in red and the red dot lines delimit the SPEED pupil diameter of the components (the useful area), their center being put at zero on the horizontal axis.}
    \label{fig:MirrorRadialProfiles}
\end{figure}

The goal of this part is to obtain the surface radial profile of the mirrors. It is on one hand to verify if the surface meets the specifications and on the other hand to build a numerical component map of the mirrors to insert it into performance simulations of the coronagraph.

A low magnification map is needed to do this measurement. Thereby data from the \textit{WYKO} were used by joining several field of views (with the minimum magnification allowed by this microscope) taken along the radial axis of the components. These radial profiles are presented in the figure~\ref{fig:MirrorRadialProfiles} (for the M1 on the left and the M2 on the right) in black, superimposed with the specification profile in red, and are taken via a radial mean over the measurements. The discrepancy between the specification and the measured profiles are found to be $63 \, nm \, rms$ and $28 \, nm \, rms$ for the M1 and the M2, respectively. The significant discrepancy seen between the red line and the data on the radial profile of the M1 (not seen on the M2) is either due to: (1) the joining technique of the data or (2) the manufacturing. As the FPM has been found in very good quality, in despite of these discrepancies we think that it comes from the joining method and that it is probably in a fair agreement with the specified $50 \, nm \, rms$.

The useful diameters of the components, prescribed to $7.7 \, mm$, are also determined here and are $7797 \pm 2 \, \mu m$ and $7731 \pm 2 \, \mu m$ for the M1 and the M2, respectively.

\subsection{Discretization discussion}
\label{sec:MDiscretization}

The radial discretization of the components is defined by two aspects: (1) the width of the gap between the rings and (2) the radial size of the smallest rings (the figure~\ref{fig:MirrorArtifacts} is an image of these two aspects). At the highest resolution, the gap between rings are measured to $3.57 \, \mu m \, rms$ on the M1 and $4.81 \, \mu m \, rms$ on the M2 and the smallest rings have a measured size of $13.56 \, \mu m \, rms$ for the M1 and $12.15 \, \mu m \, rms$ for the M2.

\begin{figure}[!h]
    \centering
    \includegraphics[width=0.5\linewidth]{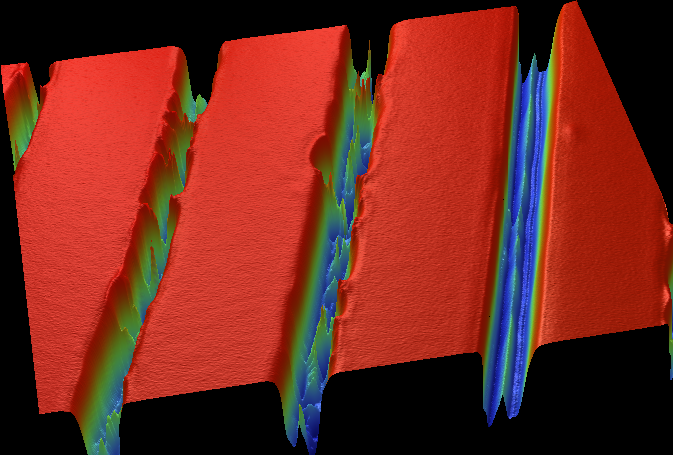}
    \caption{Image (from the \textit{WYKO} microscope) of the smallest rings (in red) on the M2 mirror at high magnification ($\times 100$). The gaps between the rings are visible in blue.}
    \label{fig:Discreization}
\end{figure}

Moreover, in the figure~\ref{fig:Discreization} we can see at the bottom of the gaps between rings some peaks of OPD. It reminds the \textit{batwings} effect (already discussed in section~\ref{sec:CosmeticAspects}) which may be seen as an interferometry measurement artifact due to the difference in depth seen in the gaps between rings and the slope of the mirror profile. This new example of such artifacts corroborates this hypothesis.

\section{Discussion and prospective}
\label{sec:Prospective}

The FPM characterization \cite{martinez2020design}, yielded a corresponding 2-D numerical map in order to model it in simulations. The as-built map of the FPM is only made from the measurements inside a defined area (see section~\ref{sec:EngravingMeasurements}), rejecting the edge effects, because the measurements was done at a low magnification ($\times 5$). These small scale effects are high-frequency noise rejected by the Lyot stop such that it is not expected to impact the coronagraphic performances.

\begin{figure}[!h]
    \centering
    \includegraphics[width=0.45\linewidth]{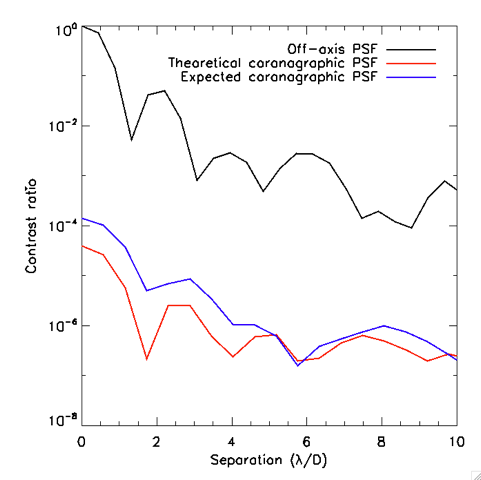}
    \caption{Contrast ratio as a function of the angular separation. In black the theoretical off-axis PSF, in red the theoretical on-axis PSF (with the coronagraph) and in blue the expected coronagraphic PSF.}
    \label{fig:CoronagraphicPerf}
\end{figure}

A coronagraphic numerical code \cite{beaulieu2018end} written in IDL including Fresnel propagation using PROPER \cite{krist2007proper} takes into account the SPEED pupil, the theoretical PIAA mirrors and Lyot stop and the as-built numerical FPM map. The wavelength is set to $1.65 \, \mu m$ (H-band), and the source is point-like. As a consequence, it simulates the coronagraphic performances of the as-built FPM. The figure~\ref{fig:CoronagraphicPerf} presents the coronagraphic contrast obtained by this simulation in blue to be compared to the specification in red.

The average contrast degradation factor (from IWA to $8 \, \lambda / D$) is $4.1$. More specifically, the contrast degradation factor at the IWA is 8.6 and at farther angular separations than $\sim 10 \, \lambda / D$ no degradation is observable. Thus the main impact in the performances is located at small angular separations (between the IWA and $5 \, \lambda / D$). These degradation are expected to be corrected with wavefront control and shaping algorithm \cite{beaulieu2018end} thanks to the high-contrast channel installed on the SPEED testbed.

The PIAACMC mirror manufacturing errors were not simulated because the joining method of several images to obtain a complete radial profile (see section~\ref{sec:RadialProfile}) of the components introduces errors in the numerical maps which make the as-built mirrors performances difficult to assess. The next step on the coronagraph components is their integration on the SPEED testbed.

\section{Conclusion}

We show the characterization of the components of the SPEED PIAACMC, previously manufactured. The cosmetic aspects are found satisfying regarding their expected degradation on the coronagraph performances. Moreover, the FPM, M1 and M2 mirrors surfaces meet the specifications with $3 \, nm \, rms$ of deviation for the FPM and on the order of $50 \, nm \, rms$ as specified for both mirrors. This characterization has also yielded numerical maps of the FPM measured surfaces, taking into account the discrepancies with the specifications but excluding the small scale effects (such as \textit{batwings} on the segment edges). This as-built map is then used in simulations to assess the impact of the manufactured FPM on the coronagraphic contrast which has been found low enough to be well corrected by wavefront control on the SPEED testbed. Eventually, the integration of these components on the SPEED experiment, soon conducted, will give the limiting contrast achievable.



 

\bibliography{report} 
\bibliographystyle{spiebib} 

\end{document}